\documentclass[aip,pop]{revtex4-1}
\usepackage[final]{graphicx}
\usepackage[pdftex]{color}

\PassOptionsToPackage{hyphens}{url}
\usepackage[hypertexnames=false]{hyperref}
\usepackage{natbib}
\usepackage{amssymb}
\usepackage{amsmath}
\usepackage{color}
\usepackage{xfrac}
\usepackage[normalem]{ulem}
\usepackage{enumitem}

\begin{document}
\newcommand{\be}{\begin{equation}}
\newcommand{\ee}{\end{equation}}
\newcommand{\bea}{\begin{eqnarray}}
\newcommand{\eea}{\end{eqnarray}}

\title{Dispersion and the Speed-Limited Particle-in-Cell Algorithm}

\author{Thomas G. Jenkins}
\email[]{tgjenkins@txcorp.com}
\homepage[]{https://nucleus.txcorp.com/~tgjenkins}
\affiliation{Tech-X Corporation, 5621 Arapahoe Avenue Suite A, Boulder, Colorado 80303, USA}

\author{Gregory R. Werner}
\affiliation{Center for Integrated Plasma Studies, University of Colorado, Boulder, Colorado 80309, USA}

\author{John R. Cary}
\affiliation{Tech-X Corporation, 5621 Arapahoe Avenue Suite A, Boulder, Colorado 80303, USA}
\affiliation{Center for Integrated Plasma Studies, University of Colorado, Boulder, Colorado 80309, USA}

\date{\today}

\begin{abstract}
This paper discusses temporally 
continuous and discrete forms of the speed-limited particle-in-cell
(SLPIC) method first treated by Werner {\it et al.}~[Phys. Plasmas {\bf 25}, 123512 (2018)].
The dispersion relation for a 1D1V electrostatic plasma whose fast particles are speed-limited
is derived and analyzed.  By examining the normal modes of this dispersion relation, 
we show that the imposed speed-limiting substantially reduces the frequency of 
fast electron plasma
oscillations while preserving the correct physics of lower-frequency plasma dynamics (e.g. ion acoustic
wave dispersion and damping).  We then demonstrate how the timestep constraints of conventional
electrostatic particle-in-cell methods are relaxed by the speed-limiting approach, thus enabling
larger timesteps and faster simulations.  
These results indicate that the SLPIC method is a fast, accurate, and powerful technique for 
modeling plasmas wherein electron kinetic behavior is nontrivial (such that a fluid/Boltzmann 
representation for electrons is inadequate) but evolution is on ion timescales.
\[ \]

This is the accepted manuscript version of the journal article whose reference is: \\
T. G. Jenkins, G. R. Werner, and J. R Cary, ``Dispersion and the speed-limited particle-in-cell algorithm'', 
Phys.~Plasmas {\bf 28}, 062107 (2021), \url{https://doi.org/10.1063/5.0046935}.

\end{abstract}


\maketitle %\maketitle must follow title, authors, abstract and \pacs

\section{Introduction}
The speed-limited particle-in-cell (SLPIC) method \cite{Werner:slpic} is a relatively new  
plasma modeling technique.  It is most suitable for discharges in which the physics of
interest occurs on relatively slow timescales (e.g., ion transport/profile relaxation) but is nevertheless 
tied to kinetic electron behaviors that a fluid/Boltzmann model cannot capture (e.g.~distribution
function modifications from neutral collisions or sheath interactions, or Landau damping).  
In such simulations, one is 
typically constrained to model both the heavy, slow ion species and the light, fast electrons using 
conventional particle-in-cell (PIC) techniques.  The ensuing computational costs can be 
(possibly unaffordably) high;
simulation timesteps must resolve the electron plasma frequency since kinetic electrons are present, but 
ion timescales of interest may exceed the electron oscillation period by many orders of magnitude. 

In the SLPIC approach, conventional PIC is modified to artificially slow down `fast' behaviors
which are numerically troublesome, despite being physically unimportant for the physics of interest.  
Larger simulation timesteps can thus be
used while retaining the detailed physics behaviors associated with the slower, longer timescales.  The
specifics of the SLPIC method will be explained in a later section of this paper, but we will note here that 
numerical experiments using SLPIC simulations to model sheath formation in an argon plasma 
have shown that remarkable speedup factors
(160 times faster than conventional PIC methods 
\footnote{Subsequent code development has enabled speedup factors of greater than 250 for this discharge, relative to conventional PIC.  Detailed discharge properties are provided in Ref. \citenum{Werner:slpic}.}) 
can be achieved \cite{Werner:slpic}.  When SLPIC can be appropriately used for modeling, 
it is both accurate and powerful.

A concept understood since the early days of PIC modeling is that while one may 
``recover more of the essence of the situation being simulated by changing the interaction laws'', 
such changes are accompanied by costs: 
``the more one meddles with the `laws' of nature the more one must understand the
consequences.'' \cite{langdon1970}  While this quote in its original context refers to the various
approximations used in PIC simulation (e.g. finite-sized particles, grid spacings, and timesteps), in this 
paper we explore its relevance to SLPIC.  Although SLPIC is in many ways similar to conventional PIC, 
the ways in which it is different introduce additional effects that one must understand in order 
to have confidence in its provided solutions -- and this is true even independent of any effects 
imparted by finite-sized particles, grid spacings, and timesteps (though such effects are not unimportant).
Fundamentally, SLPIC and PIC methods both seek to statistically approximate the evolution of smooth 
particle distribution functions in a multidimensional phase space, in response to self-consistent 
fields and forces
-- but the underlying evolution equations of the two methods are different and will yield different physics (e.g. linear plasma wave dispersion) even before any particle-based approximations are made.

In this work, therefore, we focus on developing an understanding of the behavior of a plasma evolving 
with speed-limited dynamics (hereafter SLD).  In SLD,
the plasma
is governed by continuous, `SLPIC-like' equations of motion that differ from the ones
governing plasma evolution in our universe,
but which approximate them in certain limits that we will quantify.
For purposes of comparison, we 
also designate 
the dynamics of plasma evolution in our universe as `ordinary dynamics' (OD).
SLPIC and PIC simulation methods are, respectively, the discrete numerical
analogues of the SLD and OD that we will explore.
[Alternatively, one can think of SLD or OD respectively as continuous 
limits of SLPIC or PIC, wherein 
both the timestep $\Delta t$ and the grid spacing $\Delta x$ approach zero as the velocity
dependence of the distribution function $f_{\alpha}$ becomes smooth.]
We will show that some well-understood physics processes
from OD persist in SLD,
and that other processes are substantially modified, some of them in very helpful ways.

More specifically, in this paper we derive and analyze the analytical dispersion relation 
in an electrostatic, collisionless, unmagnetized plasma evolving according to modified
Vlasov-Poisson equations 
which govern SLD.
Such a plasma will have different wave modes, with different dispersion, relative
to a real plasma evolving 
with ordinary dynamics (OD).
By comparison with the dispersive behavior that arises in the
conventional (OD) Vlasov-Poisson system, we demonstrate both analytically and numerically that 
the speed-limiting of SLD can quantifiably modify high-frequency behaviors of this plasma (electron plasma
oscillations) while leaving low-frequency motion (ion acoustic wave decay via electron Landau damping) undisturbed.  

Section \ref{sec:SLPIC} of this paper explains the SLD concept, together with its connections both to the SLPIC algorithm and to the conventional kinetic theory of OD.  In Section \ref{sec:ES},
we discuss the behavior of an electrostatic ion-electron plasma that evolves with SLD,
and derive the dispersion relation associated with this plasma.  
Section \ref{sec:analysis} contains an analysis of the various waves
permitted by this dispersion relation, together with the new behaviors
imparted by SLD
relative to known OD behaviors.  We demonstrate that
the speed-limiting significantly relaxes a fundamental numerical constraint 
associated with conventional PIC methods and makes faster numerical simulation possible.
Section \ref{sec:linearResponse} then considers the spatial fluctuation spectrum
associated with SLD and shows it to be the same as that of OD; we briefly compare SLPIC and PIC
simulations to demonstrate this point.
Finally, in Section \ref{sec:conclusion}, we summarize our findings, review additional
research directions which these findings might enable, and discuss various 
applications for the SLPIC concept in plasma modeling more generally.

\section{Phase space evolution and its connection to PIC and SLPIC methods}
\label{sec:SLPIC}
To kinetically model plasma with OD (i.e. using the familiar physics of the real world),
we first consider the self-consistent evolution of a distribution
function $\hat{f}_{\alpha}(\mathbf{x},\mathbf{v},t)$ of physical particles of species $\alpha$.
This distribution evolves according to a phase-space continuity equation
\begin{equation}
\label{eq:ConserveVlasov}
   {\partial \over \partial t} \hat{f}_{\alpha}(\mathbf{x},\mathbf{v},t)
   + {\partial \over \partial \mathbf{x}} \cdot [\mathbf{v} \hat{f}_{\alpha}(\mathbf{x},\mathbf{v},t)]
   + {\partial \over \partial \mathbf{v}} \cdot [\mathbf{a}(\mathbf{x},\mathbf{v},t) \hat{f}_{\alpha}(\mathbf{x},\mathbf{v},t)]
   = 0,
\end{equation}
where $\mathbf{a}(\mathbf{x},\mathbf{v},t)$ is the self-consistent Lorentz acceleration
experienced by a physical particle in the distribution at position $\mathbf{x}$ with velocity $\mathbf{v}$
at time $t$, in response to local (microscopic) electromagnetic fields.
For Hamiltonian
systems, the additional phase-space preserving constraint $(\partial/\partial \mathbf{x})  \cdot \mathbf{v} +
(\partial/\partial \mathbf{v}) \cdot \mathbf{a} = 0$
of Liouville's theorem allows us to rewrite Eq. (\ref{eq:ConserveVlasov}) in the
familiar Klimontovich form
\begin{equation}
\label{eq:Klimontovich}
   {\partial \over \partial t} \hat{f}_{\alpha}(\mathbf{x},\mathbf{v},t)
   + \mathbf{v} \cdot \nabla \hat{f}_{\alpha}(\mathbf{x},\mathbf{v},t)
   + \mathbf{a} \cdot {\partial \over \partial \mathbf{v}} \hat{f}_{\alpha}(\mathbf{x},\mathbf{v},t) = 0,
\end{equation}
which can be formally solved by the method of characteristics.  Along characteristic trajectories
\begin{eqnarray}
\label{eq:charTrajXKlim}
{d \mathbf{x}_{j\alpha}(t) \over dt} & = & \mathbf{v}_{j\alpha}(t) ~,\\
\label{eq:charTrajVKlim}
{d \mathbf{v}_{j\alpha}(t) \over dt} & = & \mathbf{a}[\mathbf{x}_{j\alpha}(t),\mathbf{v}_{j\alpha}(t),t] ~,
\end{eqnarray}
the value of the distribution function $\hat{f}_{\alpha}[\mathbf{x}_{j\alpha}(t), \mathbf{v}_{j\alpha}(t), t]$ is preserved; writing the distribution function in the form
\begin{equation}
\label{eq:discreteSum}
   \hat{f}_{\alpha}(\mathbf{x},\mathbf{v},t) = \sum_{j=1}^{N_{\alpha}^{p}} 
   \delta[\mathbf{x} - \mathbf{x}_{j\alpha}(t)]
   \delta[\mathbf{v} - \mathbf{v}_{j\alpha}(t)]
\end{equation}
solves Eq.~(\ref{eq:Klimontovich}) and captures the detailed microscopic behavior of each of the $N_{\alpha}^{p}$ 
particles in response to the electromagnetic fields they (and the other species in the system) produce.

For realistic physical particle counts, this microscopic behavior is far too detailed to simulate numerically
in most plasmas, and the singular nature of Eq.~(\ref{eq:discreteSum}) is likewise problematic.  If, instead, 
one passes to the continuum limit (subdividing the discrete particle charges and masses in a manner that
preserves the local volumetric charge, mass, and energy content), one arrives at the
Vlasov equation,
\begin{equation}
\label{eq:VlasovK}
   {\partial \over \partial t} f_{\alpha}(\mathbf{x},\mathbf{v},t)
   + \mathbf{v} \cdot \nabla f_{\alpha}(\mathbf{x},\mathbf{v},t)
   + \mathbf{a} \cdot {\partial \over \partial \mathbf{v}} f_{\alpha}(\mathbf{x},\mathbf{v},t) = 0,
\end{equation}
which describes the evolution of a continuous (nonsingular) phase space fluid $f_{\alpha}$ from which
the effects of particle discreteness have been removed.  More detailed discussion of this transition,
which considers ensemble averages of Eq.~(\ref{eq:Klimontovich}), two-particle and higher-order
correlation terms, collision operators\footnote{Collisional effects, which would
replace the zero on the right-hand side of Eq. (\ref{eq:Vlasov}) with source or sink terms, could
also be included in this equation; SLPIC is compatible with conventional PIC-MCC techniques for modeling
collisional plasmas.  We will not consider collisional effects in this work, but future publications
demonstrating collisional SLPIC discharges are anticipated.}, etc., has been considered by
other authors\cite{montgomery,swanson,scheiner2019}, but Eq.~(\ref{eq:VlasovK}) suffices for our purposes here.  Like Eq.~(\ref{eq:Klimontovich}),
it can also be solved by the method of characteristics; with trajectories evolving according to Eqs.~(\ref{eq:charTrajXKlim}) -- (\ref{eq:charTrajVKlim}), we may formally write
\begin{equation}
\label{eq:discreteSumVlasov}
   f_{\alpha}(\mathbf{x},\mathbf{v},t) = \sum_{j=1}^{N_{\alpha}} w_{j\alpha}(t)
   \delta[\mathbf{x} - \mathbf{x}_{j\alpha}(t)]
   \delta[\mathbf{v} - \mathbf{v}_{j\alpha}(t)]
\end{equation}
and can verify that it is a solution of Eq.~(\ref{eq:VlasovK}).
The smooth distribution is now represented as a set of $N_{\alpha}$ discrete 
macroparticles which evolve along the trajectories given by
Eqs.~(\ref{eq:charTrajXKlim} -- \ref{eq:charTrajVKlim}).  The weight function $w_{j\alpha}$ is representative 
(in some statistical sense) of
the local value of $f_{\alpha}$ in a region near the particle's initial point on the phase space trajectory.
PIC simulation techniques build upon the fundamental concept that
Eqs.~(\ref{eq:charTrajXKlim} - \ref{eq:charTrajVKlim}) and (\ref{eq:discreteSumVlasov}) solve Eq.~(\ref{eq:VlasovK}), 
and that 
a sufficiently large number $N_{\alpha}$ of macroparticles, distributed so as to adequately resolve
the relevant regions of the phase space, can statistically represent the 6D+time evolution of 
the smooth distribution function $f_{\alpha}(\mathbf{x},\mathbf{v},t)$.  

In this paper we will consider Eqs.~(\ref{eq:charTrajXKlim}) - (\ref{eq:charTrajVKlim}) and (\ref{eq:VlasovK}) as the fundamental equations describing plasma evolution under OD.

Techniques for mapping the equations of PIC onto discrete computational
timesteps and finite grids (broadening the spatial extent of macroparticles from delta-functions to
small-but-finite widths) are discussed extensively in existing literature 
\cite{BirdsallPIC,HockneyPIC,langdon1970jcp,langdon1970,okuda1970,okuda1972,langdon1979};
detailed explanations and/or derivations of such
 techniques will not be discussed here except as needed. 
For the present it suffices to note (as the previously cited
works discuss)
that finite-sized grid cells and timesteps impose a number of
constraints on conventional explicit PIC simulations:

\begin{itemize}
\item The {\it Debye length resolution constraint}, that a representative grid cell size $\Delta x$ 
should adequately
resolve the Debye length $\lambda_{D\alpha}$ associated with any of the species in the simulation 
in order to avoid numerical heating effects [the Debye length of species $\alpha$ is defined as 
$\lambda_{D\alpha}^{2} \equiv \epsilon_{0} T_{\alpha}/(q_{\alpha}^{2} n_{\alpha})$, where 
$\{q_{\alpha}, n_{\alpha}, T_{\alpha}\}$ are the species charge, density, and temperature (in units of
energy) and $\epsilon_{0}$ is the permittivity of free space];
\item The {\it cell-crossing-time constraint}, that the distance traveled by any macroparticle in
the simulation during a finite timestep $\Delta t$ should not exceed a representative grid cell size $\Delta x$,
so that forces experienced by a particle during a single simulation timestep are adequately resolved; and
\item The {\it plasma oscillation constraint}, that the plasma frequency $\omega_{p}$
constrains the timestep through the relation $\omega_{p} \Delta t \le 2$; otherwise, 
numerical instability of these high-frequency oscillations ensues [the
plasma frequency is defined as $\omega_{p}^{2} \equiv \sum_{\alpha}
\omega_{p\alpha}^{2}$, with the species plasma frequency $\omega_{p\alpha}$ defined through $\omega_{p\alpha}^{2} \equiv q_{\alpha}^{2} n_{\alpha}/(\epsilon_{0} m_{\alpha})$.  Here, $m_{\alpha}$ is the mass of species $\alpha$ and the other quantities were
defined previously].
To be precise, this constraint arises from a more general requirement that every plasma
mode frequency must be resolved by the simulation timestep.  But for a large class of problems, including the ones considered 
in this work, the highest mode frequencies are on the order of $\omega_{p}$.
\end{itemize}

These constraints can impose significant restrictions
on a plasma simulation.  Low temperatures and/or high densities decrease 
the Debye length and the allowable grid size, necessitating the use of finer grids and smaller 
timesteps.  Further, when both cold, massive ions and hot, light electrons are simulated with PIC, the 
timesteps imposed by the plasma oscillation constraint (now dominated by fast electron motion since $\omega_{p} \sim \omega_{pe}$) are so small that ions may hardly 
move at all in that time interval.  
Numerical techniques such as subcycling\cite{adam}, in which ions are pushed less frequently and
with a larger effective timestep, can provide minor computational savings, but this gains one at most a factor of $\sim2$ in speedup (for typical cases with comparable electron and ion particle counts) since 
the timestep constraints arising from electron motion remain.
For simulations where many periods of harmonic ion motion are of 
interest, the number of timesteps required can be enormous.  

The speed-limited particle-in-cell approach, and the more general speed-limiting concepts  
we explore in this work, are motivated by a desire to relax some of these constraints.  The key idea
is simple: the fastest particles and highest-frequency wave phenomena necessitate the smallest timesteps, and
if these fast particle and wave motions can be slowed, the timestep constraints can be relaxed.

Accordingly, we introduce speed-limited dynamics (SLD), 
wherein equations from the derivation of the SLPIC method presented in 
Ref. \protect \citenum {Werner:slpic} govern the plasma dynamics.  Here, a
distribution function
$f_{\alpha}$ of species $\alpha$ evolves as prescribed by a modified Vlasov equation
\begin{equation}
\label{eq:slpicVlasovNZ}
   {\partial \over \partial t} f_{\alpha}(\mathbf{x},\mathbf{v},t)
   + \beta(\mathbf{v}) \mathbf{v} \cdot {\partial \over \partial \mathbf{x}} f_{\alpha}(\mathbf{x},\mathbf{v},t)
   + \beta(\mathbf{v}) \mathbf{a} \cdot {\partial \over \partial \mathbf{v}} f_{\alpha}(\mathbf{x},\mathbf{v},t) 
   = 0 ~,
\end{equation}
wherein a speed-limiting function $\beta(\mathbf{v})$ in the range $(0,1]$ has been 
introduced.  This function
transitions from values at or near unity (for ``slow'' particles) to values approaching $v_{0}/|\mathbf{v}|$
(for ``fast'' particles), and we can understand its effect by looking at the characteristic trajectories of the modified Vlasov equation
\begin{eqnarray}
\label{eq:charTrajXslpic}
{d \mathbf{x}(t) \over dt} & = & \beta[\mathbf{v}(t)] \mathbf{v}(t) ~,\\
\label{eq:charTrajVslpic}
{d \mathbf{v}(t) \over dt} & = & \beta[\mathbf{v}(t)] 
\mathbf{a}[\mathbf{x}(t),\mathbf{v}(t),t] ~.
\end{eqnarray}
The product $|\mathbf{v}| \beta(\mathbf{v})$, the speed at which an element of phase space
changes its position $\mathbf{x}(t)$ as it moves through the phase space, now has value $\sim |\mathbf{v}|$
at low velocities but is limited to value $v_{0}$ (in the original vector direction of motion) 
at high velocities.  We will hereafter refer to $v_{0}$ as the ``speed limit''; it is the
upper bound on the rate at which motion in the position coordinate of phase space may proceed.
Accordingly, some nuance is required
in discussing the phase space evolution since the meaning of `velocity' is now ambiguous.
An element of phase space has both 
a `true velocity' (the phase space coordinate 
$\mathbf{v}$) and a `pseudo-velocity' $d\mathbf{x}/dt = \beta \mathbf{v}$
(the speed and direction at which it is permitted to move 
from one physical space coordinate to another)\footnote{}.  In a given time interval, elements
in the phase space with large true velocity $\mathbf{v}(t)$ 
[so that $\beta(\mathbf{v}) \ll 1$] experience both 
smaller pseudo-velocities [Eq.~(\ref{eq:charTrajXslpic})] as they move through the space, and 
smaller changes to these pseudo-velocities (pseudo-acceleration)
in response to applied forces [Eq.~(\ref{eq:charTrajVslpic})].
Elements in the phase space with small true velocity 
experience no speed-limiting [$\beta(\mathbf{v}) \sim 1$] and evolve in the same manner as
their OD counterparts, as in Eqs.~(\ref{eq:charTrajXKlim}) and (\ref{eq:charTrajVKlim}).
Transitions across the boundary $|\mathbf{v}| = v_{0}$ in either direction
are well-defined; this has already been demonstrated for SLPIC in Fig. 4 of Ref.~\citenum{Werner:slpic}, 
wherein particles are not observed to `pile up' at the boundary.

In this paper we will consider Eqs.~(\ref{eq:slpicVlasovNZ}) - (\ref{eq:charTrajVslpic}) as the fundamental
equations describing plasma evolution under SLD.

Solutions to the SLD Vlasov equation, Eq.~(\ref{eq:slpicVlasovNZ}), 
can be represented statistically in the same manner as outlined above, setting
\begin{equation}
\label{eq:discreteSumElec}
   f_{\alpha}(\mathbf{x},\mathbf{v},t) = \sum_{j=1}^{N_{\alpha}} w_{j\alpha}(t)
   \delta[\mathbf{x} - \mathbf{x}_{j\alpha}(t)] 
   \delta[\mathbf{v} - \mathbf{v}_{j\alpha}(t)]
\end{equation}
for a suitably large number $N_{\alpha}$ of macroparticles evolving along the characteristic
trajectories described by Eqs.~(\ref{eq:charTrajXslpic}) -- (\ref{eq:charTrajVslpic}).
This is the speed-limited particle-in-cell method we have presented in previous
work \cite{Werner:slpic}.  However, this work will not focus on PIC or SLPIC implementations 
of Eqs.~(\ref{eq:VlasovK}) or (\ref{eq:slpicVlasovNZ}).  Instead, we will consider these equations
analytically.

\section{A 1D1V electrostatic plasma model}
\label{sec:ES}
In this section we will apply the Vlasov equation of OD and the modified Vlasov equation
of SLD to 
model dispersion in an electrostatic, unmagnetized plasma with a single ion species, in one spatial 
dimension and one velocity-space dimension.  We will use the analytic forms of these equations
(foregoing for the moment any discussion of the effects of finite timesteps, grid spacings, or particle
sizes) to ensure that we understand the new physics that the imposed speed-limiting of SLD 
imparts.  Each species will use the kinetic equation
\begin{equation}
\label{eq:slpicElec1D}
   {\partial f_{\alpha}(x,v,t)\over \partial t} 
   + \beta(v) v {\partial f_{\alpha}(x,v,t)\over \partial x} 
   - \beta(v) {q_{\alpha} \over m_{\alpha}} {\partial \phi(x,t) \over \partial x} {\partial f_{\alpha}(x,v,t) \over \partial v}  = 0 ~.
\end{equation}
In SLD, we will use a speed-limiting function of form
\begin{equation}
\label{eq:1dbeta}
\beta(v) = 
-{v_{0} \over v} + \left( {v_{0} \over v} + 1 \right) H(v+v_{0}) 
+ \left({v_{0} \over v}-1 \right) H(v-v_{0}) ~,
\end{equation}
wherein $H(x)$ is the Heaviside function.  An equivalent representation for this
speed-limiting function is 
\begin{equation}
\label{eq:1dbetaOld}
\beta(v) = \left\{
\begin{array}{cc}
1 \qquad; & |v| \le v_{0} \\
v_{0}/|v| \qquad; \qquad & |v| \ge v_{0}
\end{array}
\right. ~.
\end{equation}
In OD, $\beta(v) = 1$ (the $v_{0} \rightarrow \infty$ limit of the SLD).
The species couple via the Poisson equation,
\begin{equation}
\label{eq:poisson}
{\partial^{2} \phi(x,t) \over \partial x^{2}} = - \sum_{\alpha} {q_{\alpha} \over \epsilon_{0}} \int_{-\infty}^{\infty} f_{\alpha}(x,v,t)~dv ~.
\end{equation}

Many equilibrium solutions of Eq.~(\ref{eq:slpicElec1D}) are possible.
We will choose physically reasonable solutions that are stationary Maxwellians
in each of the individual species (though we will allow the two species to have different temperatures,
and neglect both the collisional processes that have brought the individual species to their present
state and the interspecies collision processes that would further relax the system to a single temperature).
Formally, we write the distribution function of species $\alpha$ as 
\begin{equation}
f_{0\alpha}(x,v,t) = n_{0} \sqrt{ {m_{\alpha} \over 2 \pi T_{0\alpha}} } 
\exp \left( - {m_{\alpha} v^{2} \over 2 T_{0\alpha}} \right)  ~,
\end{equation}
where $n_{0}$ is a species-independent constant number density, $T_{0\alpha}$ is the constant
temperature of species $\alpha$, and $m_{\alpha}$ is the species mass.
With these equilibrium distributions, the corresponding equilibrium potential $\phi_{0}(x,t)$ is a 
constant that can be set to zero.

Linearizing Eqs.~(\ref{eq:slpicElec1D} -- \ref{eq:1dbeta}) in perturbed quantities, and Fourier
transforming from spacetime coordinates $\{x,t\}$ to wavenumber and frequency coordinates
$\{k,\omega\}$, yields the result
\begin{equation}
\label{eq:slpicElec1Dpert}
   -i \omega f_{\alpha1}(k,v,\omega) + i k \beta(v) v f_{\alpha1}(k,v,\omega) =  
   \beta(v) {q_{\alpha} \over m_{e}} i k \phi_{1}(k,\omega) {\partial f_{\alpha0}(v) \over \partial v}  ~,
\end{equation}
\begin{equation}
\label{eq:poissonpert}
- k^{2} \phi_{1}(k,\omega) = - \sum_{\alpha} {q_{\alpha} \over \epsilon_{0}} \int_{-\infty}^{\infty} f_{\alpha1}(k,v,\omega)~dv ~.
\end{equation}

We obtain, for the perturbed distribution functions,
\begin{equation}
\label{eq:fe1}
  f_{\alpha1}(k,v,\omega) =  
   {k v \beta(v) \over \omega - k v \beta(v)} 
   {q_{\alpha} \phi_{1}(k,\omega) \over T_{0\alpha}} n_{0} \sqrt{ {m_{\alpha} \over 2 \pi T_{0\alpha}} } 
\exp \left( - {m_{\alpha} v^{2} \over 2 T_{0\alpha}} \right)
\end{equation}
which we can then substitute into Eq.~(\ref{eq:poissonpert}).  The ensuing integrals will be
undefined for resonant velocities $v \beta(v) = \omega/k$;
we will implicitly stipulate that the integrals are to be evaluated using the Landau contour (traversing below 
any singularity) to retain both causality and the resonant physics.  (Formally, this can be shown
to be equivalent to the use of a Laplace transform, rather than a Fourier transform, in the time domain;
it also permits the generalization of $\omega$ to complex values.)
We obtain the integral relation
\begin{equation}
\label{eq:drint}
1 - \sum_{\alpha}
{q_{\alpha}^{2} n_{0} \over k^{2} \epsilon_{0} T_{0\alpha}} \int_{-\infty}^{\infty} 
   {k v \beta(v) \over \omega - k v \beta(v)} 
     \sqrt{ {m_{\alpha} \over 2 \pi T_{0\alpha}} } 
e^{- m_{\alpha} v^{2}/(2 T_{0\alpha})}
   ~dv = 0
\end{equation}
which describes the dispersive wave behavior of the plasma in 
both OD [where $\beta(v) = 1$] and SLD [where
Eq.~(\ref{eq:1dbetaOld}) defines $\beta(v)$].

In OD, the integral in Eq.~(\ref{eq:drint}) can be expressed in terms of the 
plasma dispersion function \cite{FriedAndConte}, defined as
\begin{equation}
\label{eq:zfunc}
Z(\zeta) = {1 \over \sqrt{\pi}} \int_{-\infty}^{\infty} {e^{-t^2} \over t-\zeta}~dt
\end{equation}
for $\mbox{Im}(\zeta)>0$ and by its analytic continuation for $\mbox{Im}(\zeta) \le 0$.
In SLD, this integral is more complicated; given our choice for $\beta(v)$, there are
both high-velocity regions of integration wherein $v\beta(v) = \pm v_{0}$, a constant 
(the speed limit), and low-velocity
regions wherein $v\beta(v) = v$.  We may represent the integrals over these various regions in terms of
the complementary error function and the incomplete plasma dispersion function.
The latter function was introduced by Franklin \cite{Franklin} and its properties have been
discussed extensively by Baalrud \cite{Baalrud:2013}; 
it takes the form [generalized from Eq.~(\ref{eq:zfunc})]
\begin{equation}
Z(\gamma,\zeta) = {1 \over \sqrt{\pi}} \int_{\gamma}^{\infty} {e^{-t^2} \over t-\zeta}~dt
\end{equation}
for $\mbox{Im}(\zeta)>0$ and by its analytic continuation for $\mbox{Im}(\zeta) \le 0$.
It will be useful to note that $Z(-\infty,\zeta) = Z(\zeta)$ and that $Z(\infty,\zeta)=0$.
Additional properties of this function are provided in Appendix \ref{appendix:expansions}.

In terms of these functions, we may rewrite Eq.~(\ref{eq:drint}) in the form
$D_{SLD}(k,\omega;v_{0}) = 0$, where we define
\begin{equation}
\label{eq:drint2}
D_{SLD}(k,\omega;v_{0}) \equiv 1 + 
\sum_{\alpha} {1 \over k^{2} \lambda_{D\alpha}^{2}} \left[
1 + {\zeta_{\alpha}^{2} \mbox{erfc}(\gamma_{\alpha}) \over \gamma_{\alpha}^2 - \zeta_{\alpha}^{2}} + \zeta_{\alpha}[Z(-\gamma_{\alpha},\zeta_{\alpha})
- Z(\gamma_{\alpha},\zeta_{\alpha})]
\right] ~.
\end{equation}
Here, we make use of the parameters $\gamma_{\alpha} \equiv v_{0}/(\sqrt{2} v_{t\alpha})$, 
a measure of the relative speed-limiting of species $\alpha$; $\zeta_{\alpha} =
\omega/(\sqrt{2} k v_{t\alpha})$, the conventional (and complex) argument of the plasma
dispersion function; $\lambda_{D\alpha}^{2} \equiv \epsilon_{0} T_{0\alpha}/(q_{i}^{2} n_{0})$, 
the Debye length of species $\alpha$; and $v_{t\alpha}^{2} \equiv T_{0\alpha}/m_{\alpha}$, the
thermal velocity for species $\alpha$.  The complementary
error function is related to the conventional error function by $\mbox{erfc}(x) \equiv 1 - \mbox{erf}(x)$.
The expression $D_{SLD}(k,\omega;v_{0}) = 0$ is the plasma dispersion relation of
SLD, and its analysis and solutions will be the topic of the following section.

Recalling that $\mbox{erfc}(\infty) = 0$
and the limits of the incomplete plasma dispersion function discussed above, we can show
that $D_{SLD}(k,\omega;\infty) = D_{OD}(k,\omega)$, where the latter function has the explicit form
\begin{equation}
\label{eq:drint2pic}
D_{OD}(k,\omega) = 1 
+ \sum_{\alpha} {1 \over k^{2} \lambda_{D\alpha}^{2}} [1 + \zeta_{\alpha} Z(\zeta_{\alpha})] ~,
\end{equation}
a standard result from elementary plasma kinetic theory \cite{GoldstonTextbook, ChenTextbook}.
This is the plasma dispersion relation of OD, against which solutions with SLD will be compared.

\section{Analysis of the dispersion relation}
\label{sec:analysis}
We now consider various limits of $D_{OD}(k,\omega)$ and $D_{SLD}(k,\omega;v_{0})$,
together with the waves that arise in these various limits.  For clarity, we will consider the
physics of OD first, and will then explore the changes which the speed-limiting
of SLD imparts to these familiar processes.

\subsection{Plasma oscillations}
\label{sec:wpewaves}
The OD dispersion relation, Eq.~(\ref{eq:drint2pic}), admits approximate analytic
solutions corresponding to cold-plasma oscillations in the $\zeta_{\alpha} \gg 1$ limit.
The asymptotic expansion of $Z(\zeta_{\alpha})$ in this limit,
\begin{equation}
Z(\zeta_{\alpha}) \sim -{1 \over \zeta_{\alpha}} 
\left(1 + {1 \over 2 \zeta_{\alpha}^{2}} + {3 \over 4 \zeta_{\alpha}^{4}} + \ldots \right) 
+ {\cal O}\left( e^{-\zeta_{\alpha}^{2}} \right)
\end{equation}
can be substituted into Eq.~(\ref{eq:drint2pic}) to obtain, at lowest order, 
\begin{equation}
\label{eq:wpe}
D_{OD}(k,\omega) = 1 - {\omega_{pe}^{2} \over \omega^{2}} - {\omega_{pi}^{2} \over \omega^{2}} = 0
\end{equation}
where $\omega_{p\alpha}^{2} = q_{i}^{2} n_{0}/(\epsilon_{0} m_{\alpha})$ is the square
of the plasma frequency of species $\alpha$.  These oscillations are dominated by 
electron motion (the ion term is order $m_{e}/m_{i}$ smaller than the electron term);
we may write the solution as $\omega^{2} = \omega_{pe}^{2}(1+m_e/m_i) \equiv \omega_{p}^{2}$.
We anticipate the prospect of significant changes to these 
oscillations in SLD, since the speed-limiting will preferentially 
modify the fast electron motion.

\begin{figure}
   \centering
   \includegraphics[width=.90\textwidth]{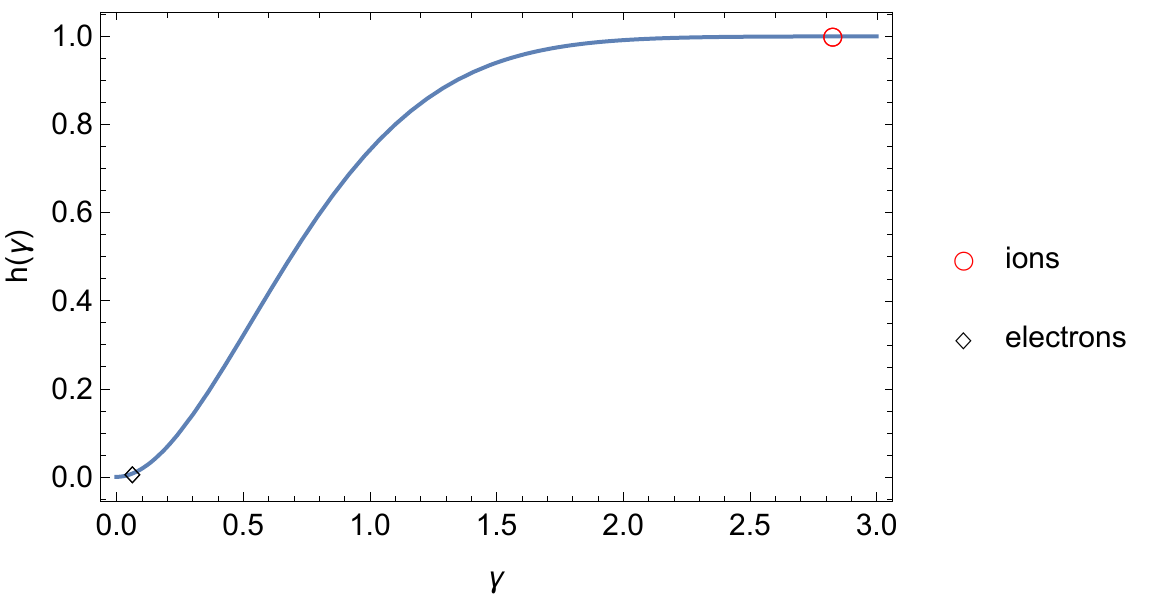}
   \caption{
   \label{hfig}
Behavior of the function $h(\gamma_{\alpha})$ in Eq.~(\ref{eq:wpeslpic}) as a function of the species
speed-limiting parameter $\gamma_{\alpha} = v_{0}/(\sqrt{2} v_{t\alpha})$.  For high values
of $v_{0}$ relative to the species thermal velocity, speed-limiting does not occur in 
densely
populated portions of the phase space and this function (a multiplicative factor in the dispersion
relation) approaches unity.  
When $v_{ti} \ll v_{0} \ll v_{te}$, considerable speed-limiting
of the electron distribution can be achieved without appreciable effect on the ion distribution, 
such that $h(\gamma_{e}) \ll 1$ while $h(\gamma_{i}) \approx 1$.  To illustrate this point, 
values of $\gamma_{\alpha}$ and the ensuing $h(\gamma_{\alpha})$ are shown for
a case with $v_{0} = 4 v_{ti}$, in a hydrogen plasma with both species at temperature 10 eV.}
\end{figure}

What is the behavior of the SLD dispersion relation, Eq.~(\ref{eq:drint2}), 
in the same $\zeta_{\alpha} \gg 1$ limit?
Using the asymptotic expansions for $Z(\gamma,\zeta)$ 
presented in Ref. \citenum{Baalrud:2013} and summarized in Appendix
\ref{appendix:expansions}, we
can show that the SLD version of Eq.~(\ref{eq:wpe}) takes the form
\begin{equation}
\label{eq:wpeslpicDR}
D_{SLD}(k,\omega;v_{0}) = 1 - \sum_{\alpha} {\omega_{p\alpha}^{2} \over \omega^{2}} 
\left( 1 + (2 \gamma_{\alpha}^{2} - 1) \mbox{erfc} (\gamma_{\alpha}) - {2 \gamma_{\alpha} e^{-\gamma_{\alpha}^{2}} \over \sqrt{\pi}} \right) = 0
\end{equation}
which reproduces Eq.~(\ref{eq:wpe}) in the $v_{0} \rightarrow \infty ~(\mbox{i.e.}~\gamma_{\alpha} \rightarrow \infty)$ limit.  It admits the solutions 
\begin{equation}
\label{eq:wpeslpic}
\omega^{2} = \sum_{\alpha} \omega_{p\alpha}^{2} h(\gamma_{\alpha}) ~~;~~
h(\gamma_{\alpha}) \equiv
\left(1 + (2 \gamma_{\alpha}^{2} - 1) \mbox{erfc} (\gamma_{\alpha}) - {2 \gamma_{\alpha} e^{-\gamma_{\alpha}^{2}} \over \sqrt{\pi}} \right) ~.
\end{equation}
The behavior of the function $h(\gamma_{\alpha})$ is shown in Figure \ref{hfig}.  
For small $\gamma_{\alpha}$, $h(\gamma_{\alpha}) \sim 2 \gamma_{\alpha}^{2}$, while
for large
$\gamma_{\alpha}$, the species is not appreciably speed-limited 
and $h(\gamma_{\alpha}) \approx 1$.  
Recalling that $\gamma_{\alpha} = v_{0}/\sqrt{2} v_{t\alpha}$, it will be instructive 
to express the argument of $h$ strictly in terms of
the electron slowing-down parameter $\gamma_{e}$; we have 
$\gamma_{i} = \gamma_{e} \sqrt{T_{e} m_{i}/(T_{i} m_{e})}$.  The presence of the ion-electron mass ratio 
suggests that unless electrons are much cooler than ions, the quantity
$\gamma_{i}$ will always be much larger than $\gamma_{e}$.  Accordingly, 
we may generally 
choose $v_{0}$ in a way that alters electron behavior but not ion behavior, consistent with
the intentional slowing down of the fastest particles in SLD while leaving slower
particles undisturbed.  
Such a $v_{0}$ will be faster than most ions but much slower than most
electrons.  For such a choice ($v_{0} = 4 v_{ti}$), Figure \ref{hfig}
also shows values of $\gamma_{e}, \gamma_{i}$, and
their corresponding $h$ values for a hydrogen plasma thermalized to 10 eV.  In this plasma, ion
motion associated with plasma oscillations does not differ appreciably between SLD
and OD, but electron motion is considerably modified.

\begin{figure}
   \centering
   \includegraphics[width=.99\textwidth]{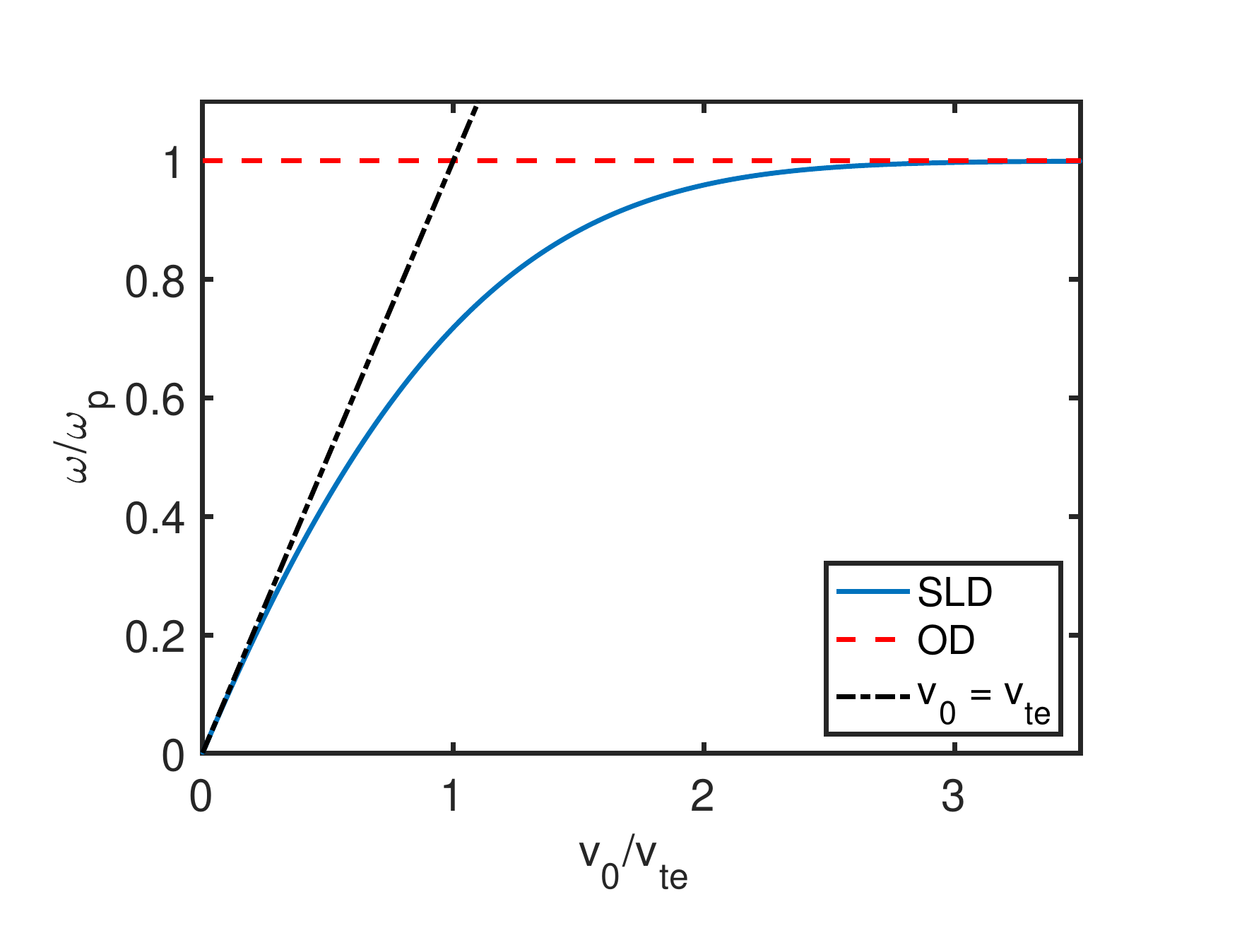}
   \caption{
   \label{fig:wpeplot}
Frequency of electron-dominated cold plasma oscillations in SLD and OD 
as a function of the 
ratio of speed limit to electron thermal speed ($v_{0}/v_{te} = \sqrt{2} \gamma_{e}$).
The electron-dominated cold plasma oscillation frequency is constant in OD (red dashed line), 
but in SLD it unphysically decreases (blue solid line) as $v_{0}/v_{te}$ is reduced.
This unphysical frequency means that plasma oscillations are not correctly
simulated, but this can greatly speed up simulation when plasma oscillations are unimportant
to the physics of interest.  When $v_{0} \ll v_{te}$, $\omega/\omega_{p} \sim v_{0}/v_{te}$.
}
\end{figure}

Since it is primarily the speed-limiting of electrons that affects the oscillation
frequencies predicted by Eq.~(\ref{eq:wpeslpic}), we examine the behavior
of these modified electron plasma oscillations
as a function of the ratio $v_{0}/v_{te}$ (providing intuition as to which velocities in a typical 
electron distribution, e.g. a Maxwellian, are restricted by the speed-limiting).
The normalized oscillation frequency of Eq.~(\ref{eq:wpeslpic}) 
is shown in Fig.~(\ref{fig:wpeplot}) for a monatomic 
hydrogen plasma with equal electron and ion temperatures.  
While the oscillation frequencies of SLD and OD are identical at 
large values of $v_{0}/v_{te}$ (minimal speed-limiting), 
reducing this ratio, and hence increasing the corresponding fraction $\mbox{erfc}(\gamma_{e})$ 
of speed-limited electrons, reduces the SLD frequency monotonically.
As $v_{0}/v_{te} \rightarrow 0$, 
the linear approximation to the plasma
frequency approaches the heuristic estimate made in Ref.~\citenum{Werner:slpic}: 
$\omega/\omega_{p} \sim v_{0}/v_{te}$.  The 
high-frequency oscillations of OD have been mapped to lower-frequency oscillations in 
SLD by the speed-limiting of electrons.

% gets made from wpeplot.m in the ./figures subdirectory

We conclude that in SLD, the frequency of conventional electron-dominated
plasma oscillations is reduced relative to OD.  In Sec.~\ref{sec:wpeconstraint} we will 
consider how this reduction relaxes 
the `plasma oscillation constraint' referred to in Sec.~{\ref{sec:SLPIC}.  But we must
first determine whether the speed-limited dynamics preserves physics associated with 
lower-frequency waves.

\subsection{Ion acoustic waves}
\label{sec:IAW}
The OD dispersion relation, Eq.~(\ref{eq:drint2pic}), also admits approximate analytic
solutions corresponding to ion acoustic waves.  Physically, these solutions are associated
with `hot' electrons ($\zeta_{e} \ll 1$) and `cold' ions ($\zeta_{i} \gg 1$), and in these limits 
the dispersion relation can be written in the approximate form
\begin{equation}
\label{eq:ionacousticDR}
k^{2} \lambda_{De}^{2} D_{OD}(k,\omega) = 1 + k^{2} \lambda_{De}^{2}
+ \zeta_{e} i \sqrt{\pi}
-  {k^{2} c_{s}^{2} \over \omega^{2}} = 0~, 
\end{equation}
where the multiplicative factor of $k^{2} \lambda_{De}^{2}$ simplifies the
algebra and where
$c_{s}$, the sound speed, satisfies the relation $c_{s}^{2} = T_{0e}/m_{i}$.  
Assuming that the imaginary part of the (now complex) $\omega = \omega_{r} 
+ i \omega_{i}$ is small, we can perform a Taylor expansion,

\begin{equation}
D(k,\omega) \approx D(k,\omega_{r}) + i \omega_{i} 
%\left(
\left.
{\partial D(k,\omega) \over \partial \omega}
\right|_{\omega = \omega_{r}}
\end{equation}
to show that this
equation permits damped wavelike solutions of the form
\begin{equation}
\label{iawdr}
\omega = \pm {k c_{s} \over \sqrt{1 + k^2 \lambda_{De}^{2}}} 
- i \sqrt{ {\epsilon \pi \over 8}} {k c_{s} \over (1 + k^2 \lambda_{De}^{2})^{2} }
\end{equation}
where $\epsilon$ is again the electron/ion mass ratio.
These are the ion acoustic wave (IAW)
modes, which are essentially longitudinal compressions of the ion mass density
that decay via Landau damping.

What is the corresponding behavior of the SLD dispersion relation?
In the $\zeta_{i} \gg 1, \zeta_{e} \ll 1$ limit, Eq.~(\ref{eq:drint2}) can be written 
with the same multiplicative factor in the form
\begin{equation}
\label{eq:ionacousticSLPICDR}
k^{2} \lambda_{De}^{2} D_{SLD}(k,\omega;v_{0}) = 1 + k^{2} \lambda_{De}^{2} -  {k^{2} c_{s}^{2} \over \omega^{2}} h(\gamma_{i})
+  i \sqrt{\pi} \zeta_{e} \left[ 
H(\zeta_{e} - \gamma_{e}) - H(\zeta_{e} + \gamma_{e})
\right] = 0~.
\end{equation}
Repeating the Taylor expansion procedure above, we can show that the
ion acoustic wave dispersion relation of SLD has approximate analytic solutions
\begin{equation}
\label{slpsiaw}
\omega = \pm \omega_{r}
%{k c_{s} \sqrt{h(\gamma_{i})} \over \sqrt{1 + k^{2}\lambda_{De}^{2}}}
- i \sqrt{{\epsilon \pi \over 8}} {k c_{s} h(\gamma_{i}) \over (1 + k^{2} \lambda_{De}^{2})^{2}}
\left[H \left( {\omega_{r} \over k} + v_{0} \right) - H \left( {\omega_{r} \over k} - v_{0} \right) \right] ~,
\end{equation}
wherein
\begin{equation}
\omega_{r} \equiv {k c_{s} \sqrt{h(\gamma_{i})} \over \sqrt{1 + k^{2}\lambda_{De}^{2}}} ~.
\end{equation}
This is the SLD equivalent to Eq.~(\ref{iawdr}).  The OD and SLD forms are 
approximately equivalent
provided that $\gamma_{i}$ is sufficiently large (so that $h(\gamma_{i}) \sim 1$, see Fig. \ref{hfig}) and
$|\omega_{r}/k| < v_{0}$ (so that the speed limit $v_{0}$ exceeds the phase velocity of the ion acoustic wave).  Since the approximation
$h(\gamma_{i}) \approx 1$ holds to one part in $10^4$ or better for $v_{0} > 4 v_{ti}$, we will
write these constraints as a single condition
\begin{equation}
v_{0} > \mbox{max}(4 v_{ti}, |\omega_{r}/k|) ~.
\label{v0restriction}
\end{equation}
When $v_{0}$ is chosen
to satisfy this condition, and when the other conditions we assumed in the derivation
[$\zeta_{i} \gg 1, \zeta_{e} \ll 1, \mbox{Im}(\omega) \ll \mbox{Re}(\omega)$] are
valid, the effects of SLD on the propagation and damping of the IAW are minimal.

What happens when the phase velocity condition is violated,
such that $v_{0} < |\omega_{r}/k|$?  In this case, the explicitly imaginary terms
in Eq.~(\ref{eq:ionacousticSLPICDR}) (proportional to Heaviside functions) vanish, 
and the solutions admitted now only capture
the real part of Eq.~(\ref{slpsiaw}).  While the IAW still propagates, its 
Landau damping is not correctly 
modeled.  This is consistent with a result that we have demonstrated in previous 
work \cite{Werner:slpic}, namely, that the correct dynamics of resonant wave-particle interactions cannot
be captured by speed-limited particle-in-cell simulations 
when particles whose velocities were previously synchronous with the wave
velocity are speed-limited.   
Particles whose speed-limiting renders them unable to keep up with the wave
cannot exchange energy with it, so the dissipative effects which lead to wave damping are
effectively turned off.
Although nothing in principle prevents us from choosing a smaller $v_{0}$ value that violates 
Eq.~(\ref{v0restriction}), the
inherent advantage of the velocity-dependent speed-limiting approach (preservation of IAW
physics) would be lost in doing so.

The analytic form of the IAW above is approximate.  The dispersion relation can
also be solved numerically to find the complex $\omega$ associated with a given $k$, 
and we have done so for a plasma with density 
$n_{0} = 5.0 \times 10^{16}~\mbox{m}^{-3}$, $T_{e0} = 10$~eV, and
$T_{i0} = 1/40$~eV.  In the supplemental materials for Ref.~\citenum{Baalrud:2013},
Matlab algorithms for evaluating $Z(\zeta)$, $Z(\gamma,\zeta)$, and their derivatives
have been provided.  We have made use of these algorithms and Matlab minimization
routines to compute $\{\omega,k\}$ 
values which satisfy the OD and SLD dispersion relations within a given tolerance,
according to norm-minimization criteria
\begin{eqnarray}
\sqrt{|D_{SLD}(k,\omega;v_{0})|^{2}} & \le & \delta_{SLD} ~, \\
\sqrt{|D_{OD}(k,\omega)|^{2}} & \le & \delta_{OD}  ~.
\end{eqnarray}
Here, the tolerance parameter $\delta \ll 1$ is a small positive number 
which constrains the allowable error in the numerical solution of a particular dispersion relation.
In these computations we fix $k$
(and, for SLD, $v_{0}$), and then numerically evaluate the function $D$ for various complex 
values of $\omega$.  Exact solutions to the dispersion relation have $D=0$; we vary $\omega$ 
in the complex plane in a manner that seeks to minimize the norm of $D$ 
and thus to approach these exact solutions.
For $\omega$ values near an exact solution this norm can in principle 
be reduced to be no greater than $\delta$.

Various values for the speed-limiting parameter $v_{0}$ can
be chosen to assess the effect of speed-limiting on IAW behavior.
Depending on the value of $v_{0}$ and $k$, to obtain numerical convergence it is sometimes 
necessary to raise $\delta_{SLD}$ to values as high as $10^{-3}$, but solutions 
can usually be found for $\delta_{OD} < 10^{-6}$ and $\delta_{SLD} <  10^{-5}$ 
(and often for $\delta$ values that are several orders of magnitude smaller, when
$k \gg 1$).

\begin{figure}
   \centering
   \includegraphics[trim=0 5.0cm 0 7cm,width=.99\textwidth]{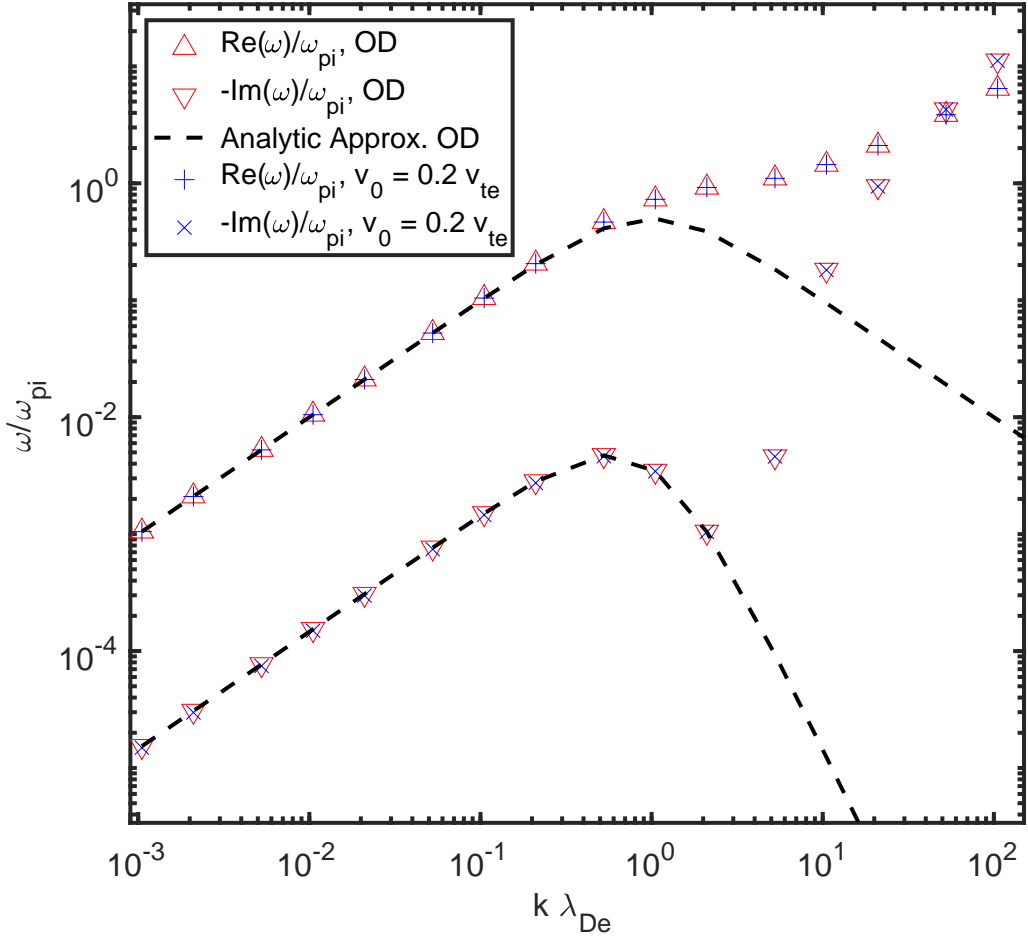}
   \caption{
   \label{fig:iawHighGamma}
Real and negative-imaginary parts of the normalized IAW frequency, computed
numerically from the OD (red) and SLD (blue)
dispersion relations, as a function of normalized wavenumber.  
Values from the approximate analytic expression, Eq.~(\ref{iawdr}), are also shown
(dashed black curves) and agree well with the exact solutions when
$k \lambda_{De} \lesssim 1$.
In the SLD case the speed limit $v_{0} = 0.2 v_{te}$.  
Even though all electrons are restricted to move through
the domain with speeds no greater than $v_{0}$, the frequency (good to within 0.5\%)
and damping rate (good to within 3\%) for the ion acoustic wave do not differ appreciably
from the OD values.  
Good agreement between SLD and OD is maintained even in the
$k \lambda_{De} > 1$ regime, where the analytic approximations underlying 
Eq.~(\ref{iawdr}) are no longer valid.
}
\end{figure}

In Fig.~\ref{fig:iawHighGamma}, we show the real and negative-imaginary parts of the
computed frequency $\omega$ for various values of wavenumber $k$ spanning several 
orders of magnitude.  In this figure, the speed limit $v_{0}$
has the value $0.2 v_{te}$; 
the analytic approximation to the dispersion relation (valid for 
$k \lambda_{De} \lesssim 1$) is also shown.
The speed-limiting does not affect the IAW behavior appreciably;
for modes whose wavelengths are large compared to the Debye length (i.e. where
the analytic approximation is valid) the SLD real frequency $\omega$ is about half a 
percent low relative to the OD value
and the SLD damping rate also drops by about 3\%.  
For shorter-wavelength modes, the effect of the speed-limiting on both the 
real frequencies and damping rates is negligible.
Nevertheless, the frequency of the modified 
plasma oscillations [from Fig.~(\ref{fig:wpeplot})] is decreased to about 
$0.2 \omega_{pe}$ for this case.

\begin{figure}
   \centering
   \includegraphics[trim=0 5.0cm 0 7cm,width=.99\textwidth]{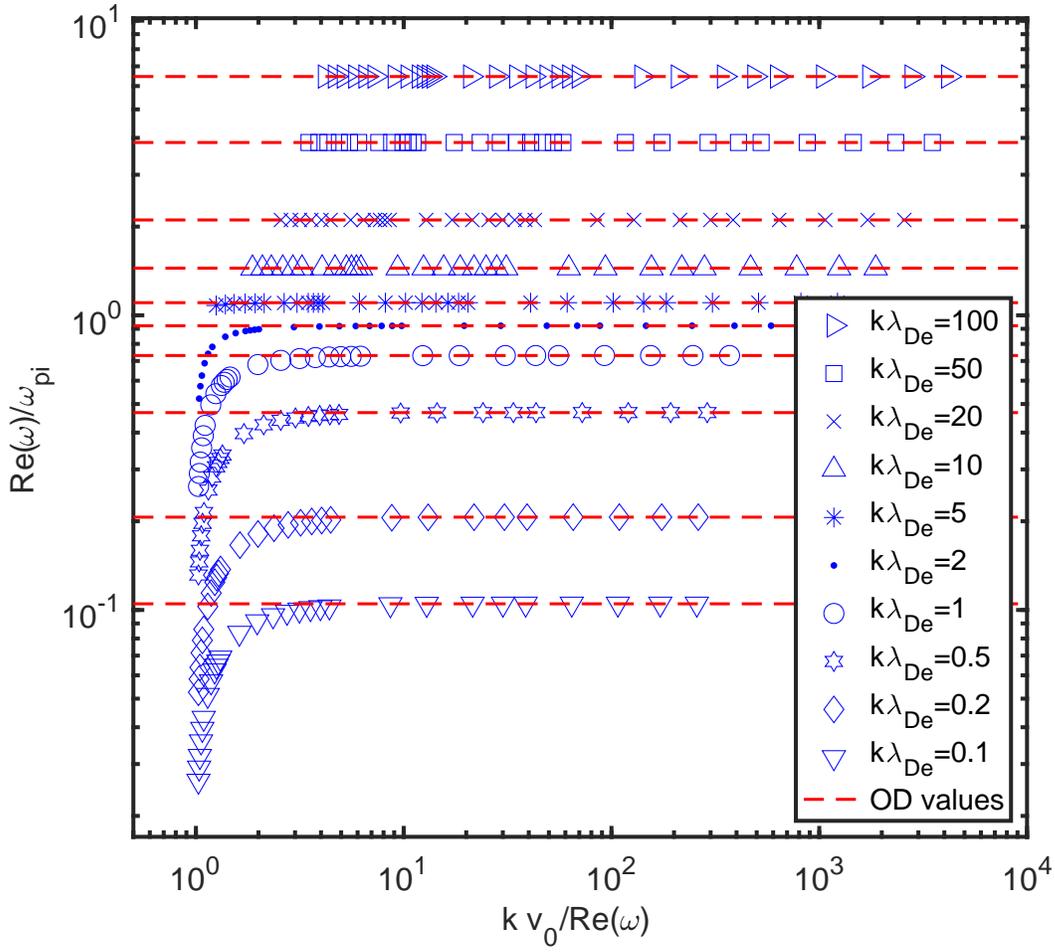}
   \caption{
   \label{fig:iawRealw2}
Variation of the normalized real part of the frequencies which solve the ion acoustic branch of
the SLD dispersion relation, 
as a function of the speed limit $v_{0}$ normalized to the wave phase velocity $v_{\phi} = \mbox{Re}(\omega)/k$, 
for various values of $k \lambda_{De}$.  
The OD frequencies, which are independent of $v_{0}$, are also shown for each $k \lambda_{De}$ value (red dashed lines).  
Although speed-limiting does not affect the mode frequencies for $v_{0}/v_{\phi} \gg 1$, 
modes whose wavelengths are large compared to the Debye length ($k \lambda_{De} \lesssim 1.0$) 
are reduced in frequency as the speed limit is reduced (moving to the left on the graph) to be of the same
order as the phase velocity.
This frequency variation is minimal for $v_{0} > 5 v_{\phi}$ and also for
modes whose wavelengths are short relative to the Debye length.  
For this case $v_{ti}/v_{te} = 0.05 \sqrt{m_e/m_i}$.}
\end{figure}

\begin{figure}
   \centering
   \includegraphics[trim=0 5.0cm 0 7cm,width=.99\textwidth]{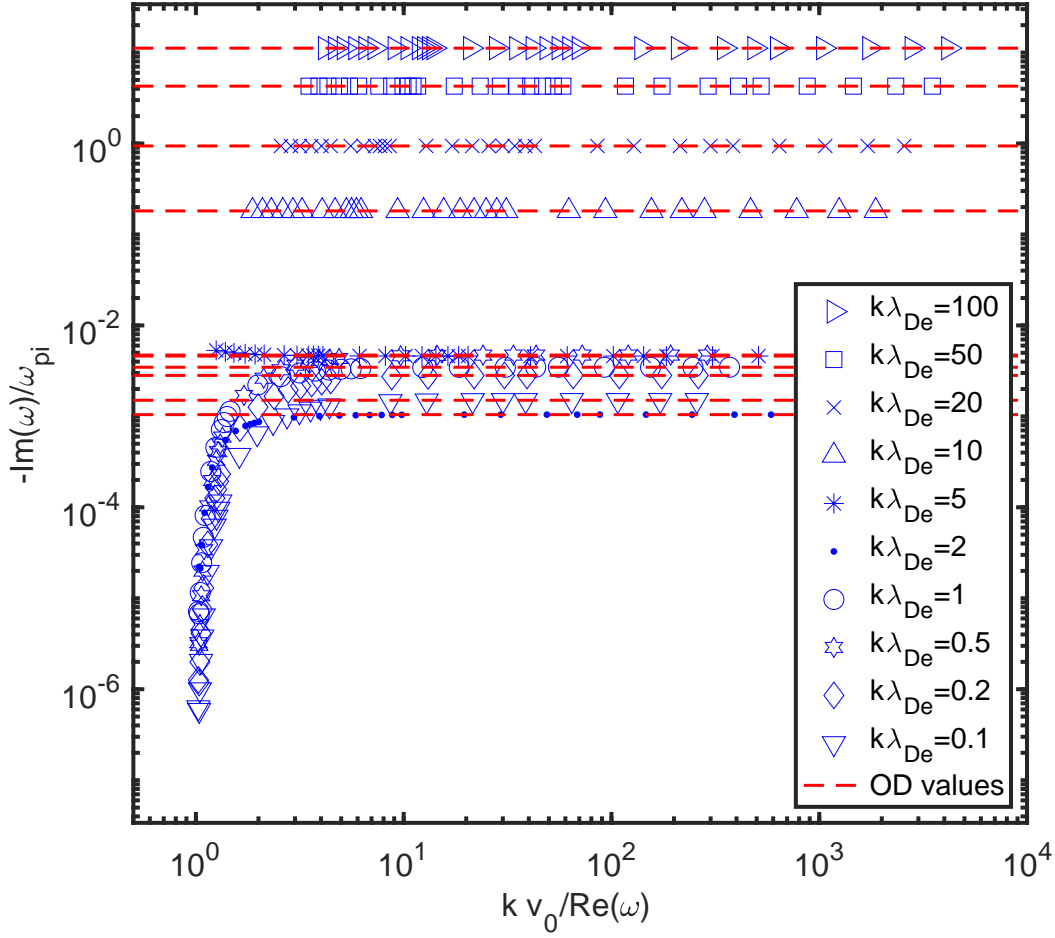}
   \caption{
   \label{fig:iawImagw2}
Variation of the normalized negative-imaginary part (damping rate) 
of the frequencies which solve the ion acoustic branch of
the SLD dispersion relation, 
as a function of the speed-limit $v_{0}$ normalized to to the wave phase velocity $v_{\phi} = \mbox{Re}(\omega)/k$,
for various values of $k \lambda_{De}$.  
The OD damping rates, which are independent of $v_{0}$, are also shown for each $k \lambda_{De}$ value (red dashed lines).
Although speed-limiting does not affect the damping rates when $v_{0}/v_{\phi} \gg 1$, 
damping rates for modes whose wavelengths are large compared to the Debye length ($k \lambda_{De} \lesssim 1.0$) 
may be reduced by several orders of magnitude as the speed limit is reduced (moving to the left on the graph)
to be of the same order as the phase velocity.
As is also true for the real frequencies (Fig.~\ref{fig:iawRealw2}), the 
damping rate variation is minimal for $v_{0} > 5 v_{\phi}$ and also for
modes whose wavelengths are short relative to the Debye length.  
For this case $v_{ti}/v_{te} = 0.05 \sqrt{m_e/m_i}$.}
\end{figure}

For more restrictive speed limits $v_{0}$, greater deviation of IAW frequencies and damping 
rates from their non-speed limited (OD) values is observed,
though still only for modes whose wavelengths are long compared to the Debye length.  
Generally speaking, real frequencies are shifted downward
as the speed limit is decreased, though only by a few percent ($<$9\% for $v_{0} = 0.05 v_{te} = 2.1 c_{s}$, 
and $<$2\% for $v_{0} = 0.1 v_{te} = 4.2 c_{s}$).  
Damping rates also (generally) decrease in magnitude as the speed limit is decreased, but the magnitude of the 
relative decrease is larger ($<$42\% for $v_{0} = 0.05 v_{te} = 2.1 c_{s}$, $<$13\% for $v_{0} = 0.1 v_{te} = 4.2 c_{s}$).  
In Figures \ref{fig:iawRealw2} and \ref{fig:iawImagw2} we have
plotted the variation of the IAW real frequency and damping rate of SLD as a function of the speed-limiting velocity
normalized to the wave phase velocity, so as to more generally quantify the effect of the speed-limiting on the IAW dispersive behavior.
These figures illustrate that lowering the speed limit has relatively little effect on either
the real IAW oscillation frequency or damping rate as long as the speed limit is higher than the wave phase velocity.
For longer-wavelength modes $(k \lambda_{De} \lesssim 1)$ in SLD, speed limits that approach
the wave phase velocity generally give rise to modest reductions in the mode frequency
(Fig.~\ref{fig:iawRealw2}) and more pronounced reductions of the IAW damping rate (Fig.~\ref{fig:iawImagw2}).
This failure to properly capture the Landau damping of the IAW arises because portions of the distribution
function that resonate with the wave (in OD) are prevented from doing so by the imposed speed-limiting
\cite{Werner:slpic}.
As we transition to shorter-wavelength modes (moving up the graph legend), absolute damping rates are increased to become comparable to the real mode frequency, while phase velocities are reduced to be of order $v_{ti}$.  
For these waves, speed-limiting does not appreciably influence the dynamics since the smallest sensible speed 
limit ($v_{0} \sim$ $4 v_{ti}$, so as not to speed-limit the bulk ion distribution) 
still exceeds $v_{\phi} = \omega/k$ for large $k$.

Accordingly, we may assert that SLD preserves the physics of IAW propagation and damping 
provided that 
the condition in Eq.~(\ref{v0restriction}) holds, namely, when ions are not speed-limited and when the speed limit reasonably exceeds the phase velocity of the IAWs in the system.
At the same time, this speed limiting considerably
reduces the frequency of plasma oscillations (as was shown in Section \ref{sec:wpewaves}).

\subsection{Normal modes and the plasma oscillation constraint}
\label{sec:wpeconstraint}

Having explored the normal modes of our 1D1V plasma, we now revisit the numerical
constraints which the use of a finite timestep $\Delta t$ will impose on particle-in-cell simulations
of this plasma.

We have noted in Section \ref{sec:SLPIC} that numerical instability will ensue in a PIC simulation
if the frequency 
of any plasma mode is not resolved, and that in particular, we must 
resolve the frequency
associated with electron plasma oscillations.  These oscillations are generally the 
highest-frequency modes in a PIC simulation containing electrons 
(because ions introduce only small corrections to the approximation 
$\omega_{p} \approx \omega_{pe}$ when the electron-ion mass ratio
is small).  Accordingly, any numerical
simulation method satisfying the constraint $\omega_{p} \Delta t \ge 2$ will resolve 
both these plasma oscillations and all other modes of lower frequency.

What is the effect of the speed limiting on this constraint?

Because of the speed-limiting, the frequency that we must resolve is now not $\omega_{p}$; rather, it is the fastest oscillation 
frequency that the speed-limiting permits.  But we have shown in Sec.~\ref{sec:analysis} that ordinarily `fast' plasma oscillations 
[see Eq.~(\ref{eq:wpeslpic})] are considerably slowed
in SLD.  Thus, in its most general
form, SLPIC replaces the plasma oscillation constraint by the result [obtained by
substituting the frequency derived in Eq.~(\ref{eq:wpeslpic}) for $\omega_{p}$]
\be
\label{eq:fac}
\Delta t \le { 2 \over 
\sqrt{\omega_{pe}^{2} h(\gamma_{e}) +\omega_{pi}^{2} h(\gamma_{i})} } \approx
{2 \over \omega_{pe} \sqrt{h(\gamma_{e})}}
\ee 
where $\gamma_{\alpha} = v_{0}/\sqrt{2} v_{t\alpha}$ is the slowing-down parameter.  
[For example, when $v_{0}/v_{te} = 0.1$, 
we have $\gamma_{e} = 0.07$, $h(\gamma_{e}) \approx
9.5 \times 10^{-3}$, and $1/\sqrt{h(\gamma_{e})} \approx 10$, thus relaxing the constraint
tenfold with minimal effect on the ion modes (assuming IAW phase velocities are low compared to $v_{0}$).]
This constraint is less restrictive than the PIC result
$\Delta t \le 2/\omega_{pe}$, and permits larger timesteps to be taken in
SLPIC simulations without instability or loss of accuracy in the low-frequency plasma modes.
It can also be shown that by limiting the maximum speed to $v_{0}$, SLPIC trivially relaxes 
the cell-crossing-time constraint by nearly the same factor (see Appendix \ref{appendix:constraints}).
Both the plasma oscillation constraint and the cell-crossing-time constraint are thus modified by 
speed-limiting to restrict $\Delta t \sim \Delta x/v_{0}$.

\section{Analysis of the fluctuation spectrum}
\label{sec:linearResponse}
From the fluctuation-dissipation theorem and the theory of linear response, the fluctuation
spectrum of an electrostatic 1D plasma in thermal equilibrium can be shown\cite{langdon1979} to take the form
\begin{equation}
\label{spect}
{\epsilon_{0} \langle E^{2} \rangle(k) \over T_{0} } = {1 \over D(k,\omega = \infty)} - {1 \over D(k, \omega = 0)} ~,
\end{equation}
\noindent
where $T_0$ is the equilibrium temperature, $\langle E^{2} \rangle(k)$ is the time-average of the
continuous spatial Fourier transform of the square of the electric field, and $D(k,\omega)$ is the dispersion
relation.
In OD, taking these limits of Eq.~(\ref{eq:drint2pic}) yields the result
\begin{equation}
\label{odspect}
\left({\epsilon_{0} \langle E^{2} \rangle(k) \over T_{0} }\right)_{OD} = {1 \over 1} - {1 \over 1 + \sum_{\alpha} {1 \over k^2 \lambda_{D\alpha}^{2}}} = {1 \over 1 + k^{2} \lambda_{D}^{2}} ~,
\end{equation}
wherein $\lambda_{D}^{2} \equiv \epsilon_{0} T_{0}/\sum_{\alpha} (q_{\alpha}^{2} n_{\alpha})$ is the
square of the plasma Debye length.

What is the behavior of the SLD fluctuation spectrum?  At low frequencies, where speed-limiting
is not expected to influence the wave dynamics, we likewise recover the same result as for OD:
\begin{equation}
D_{SLD}(k,\omega=0;v_{0}) = 1 + \sum_{\alpha} {1 \over k^{2} \lambda_{D\alpha}^{2}} ~.
\end{equation}
At high frequencies, we have shown that the behavior of solutions to the dispersion relation is significantly
altered by the speed-limiting.  Nevertheless, terms associated with speed-limiting vanish in the high-frequency limit of
the dispersion relation
\begin{equation}
D_{SLD}(k,\omega=\infty;v_{0}) = 1 + \sum_{\alpha} {1 \over k^{2} \lambda_{D\alpha}^{2}}
\left(
1-\mbox{erfc}(\gamma_{\alpha}) - {\mbox{erfc}(-\gamma_{\alpha}) \over 2} + {\mbox{erfc}(\gamma_{\alpha}) \over 2}
\right) = 1
\end{equation}
[because $\mbox{erfc}(-x) \equiv 2 - \mbox{erfc}(x)$];
the effects of speed-limiting should therefore have no bearing on the spatial fluctuation
spectrum, Eq.~(\ref{spect}).  We recover the same result as Eq.~(\ref{odspect}) for SLD,
\begin{equation}
\label{sldspect}
\left({\epsilon_{0} \langle E^{2} \rangle(k) \over T_{0} }\right)_{SLD} = {1 \over 1 + k^{2} \lambda_{D}^{2}} ~.
\end{equation}
In a PIC or SLPIC simulation, the fluctuation spectra of Eqs.~(\ref{odspect}) and (\ref{sldspect}) are altered
by finite particle size (associated with the transfer of charge and force fields between continuous particle positions and the discrete grid) as well
as by finite grid spacing (associated with the wavenumber spectrum that is able to be resolved by the
simulation).  
In addition, the introduction of a discrete grid and a finite volume leads to discrete, finite Fourier spectra and
introduces the possibility of aliasing between gridded fields and subgrid particle modes.
While we do not propose to 
discuss the effects of finite particle and grid size in detail in this work, we have used PIC and SLPIC to simulate 
a 1D single-ion-species hydrogen plasma in thermal equilibrium and have measured its fluctuation spectrum.  
For a discrete Fourier mode this spectrum satisfies a relation of the general form \cite{langdon1970jcp,okuda1972,langdon1979}
\begin{equation}
{\epsilon_{0} |E_{l}|^{2} \over n_{0} T_{0} } = {1 \over N_{p}} \left( {1 \over 1 + K^{2} \lambda_{D}^{2}/|S(k)|^{2}} \right)
\end{equation}
wherein $N_{p}$ is the number of simulation macroparticles of either species, $|S(k)|^{2}$ is a geometric 
factor associated with the particle shape, $k = 2 \pi l/L$ defines the discrete
mode index $l$, $K = k~\mbox{sinc}(k \Delta/2)$ captures the effect of the 1D Laplacian operator on the
discrete 1D grid with spacing $\Delta$ [with $\mbox{sinc}(x) \equiv \sin(x)/x$], and $|E_{l}|^{2}$ is the time-averaged norm of the $l$-th discrete mode in the 
Fourier transform of the electric field.

We modeled this scenario with the
VSim\cite{vorpal} code, using both PIC with a small timestep ($\Delta t = 1.0 \times 10^{-13}$ s) 
and SLPIC with a (50X) larger timestep.  We used a highly resolved grid (80 cells per Debye
length, with a 1D simulation length $L = 10$ Debye lengths), with equilibrium plasma
density $n_{0} = 5.0 \times 10^{16}~\mbox{m}^{-3}$
and temperature $T_{0} = 10$ eV.  100 particles per cell of each species were used; the speed limit $v_{0}$ for the
SLPIC simulations was set to one-half the electron thermal velocity.  Particles were mapped to the grid with
a three-cell (four-gridpoint) stencil using the method prescribed by Esirkepov \cite{esirkepov}.  For this mapping,
$|S(k)|^{2} = ~\mbox{sinc}^{8}(k \Delta/2)$.
\begin{figure}
   \centering
   \includegraphics[width=.99\textwidth]{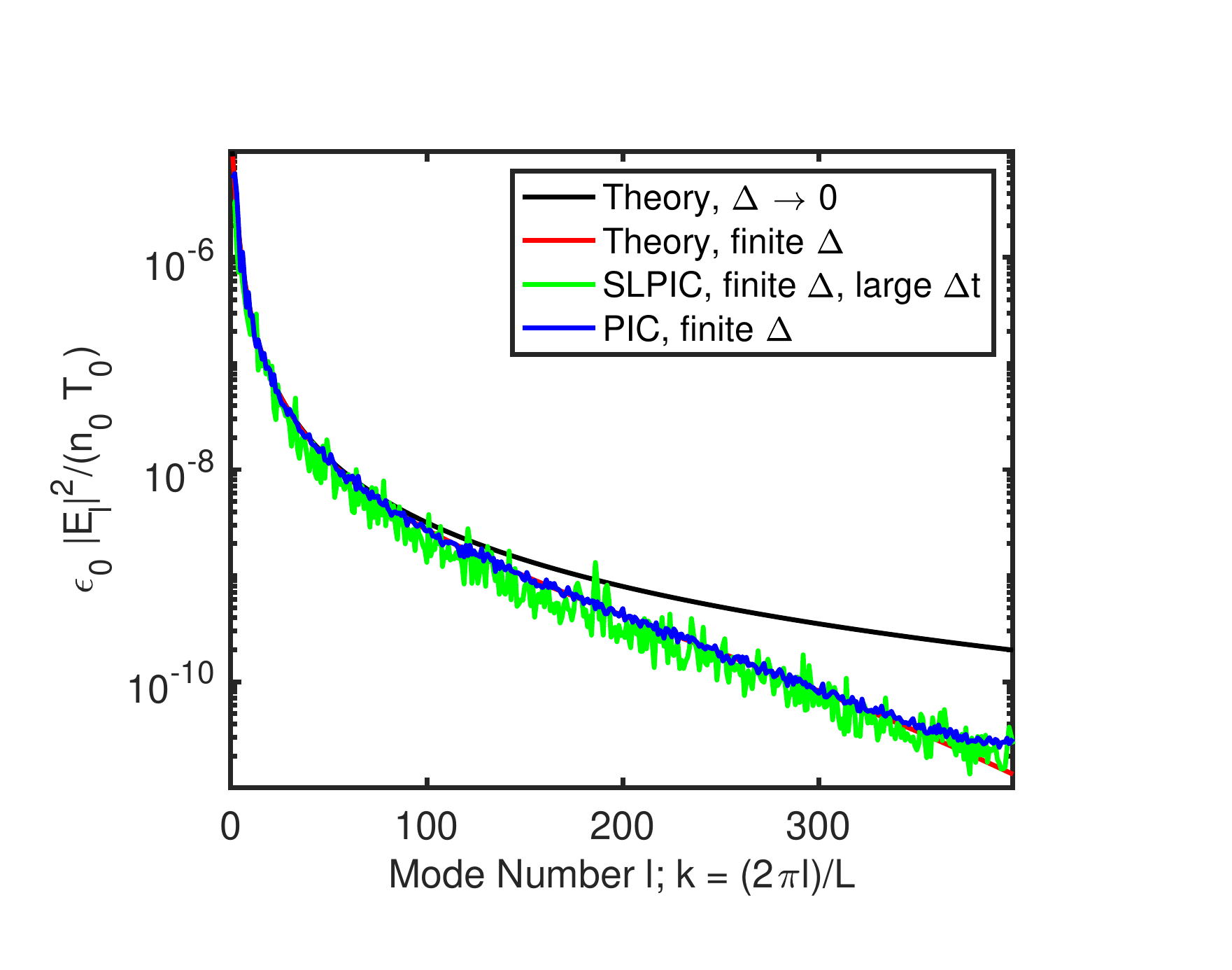}
   \caption{
   \label{spectrumfig}
Predicted fluctuation spectrum in the limit of infinite grid resolution and zero particle size (black), 
together with the
predicted behavior for finite grid resolution and particle shape (red -- and covered by the PIC data except at very high mode numbers), and the observed spectra from
SLPIC (green) and PIC (blue) simulations.  While the signals vary more with $k$ for SLPIC than for PIC, both
methods agree closely with the theory for finite $\Delta$.}
\end{figure}
As shown in Figure \ref{spectrumfig}, both PIC and SLPIC capture the general behavior of the fluctuation spectrum 
when the effects of finite grid size and particle width are accounted for.  
Larger variation between adjacent modes in the spectrum is observed for SLPIC relative to PIC,
an effect which becomes more prominent as $\gamma_{e}$ is decreased (stronger speed-limiting)
and which is perhaps a function of interparticle correlations imposed by the speed-limiting constraint (since 
all fast particles now move with the same pseudo-velocity in the domain). 
Nevertheless, both methods agree closely with the theory.  SLPIC simulations with smaller timesteps
(identical to PIC) were not seen to differ substantially from the large-timestep result (green curve) in the
figure.

A more detailed consideration of the role of finite grid spacing and 
particle width in SLPIC is a topic of ongoing interest, and we anticipate future efforts along these lines.

\section{Conclusions}
\label{sec:conclusion}
In this paper we have discussed the linear wave dispersion 
in a 1D1V unmagnetized electrostatic
plasma that evolves with both ordinary (OD) and speed-limited (SLD) dynamics.
We have demonstrated that speed-limiting can effectively
reduce the frequency of fast electron oscillations
while quantitatively preserving low-frequency
ion and electron motion, e.g.~the physics needed to correctly model the Landau damping
of ion acoustic waves.  We have also shown that this speed-limiting relaxes the
``plasma oscillation constraint'' of the conventional PIC method, permitting
larger timesteps, and have demonstrated that the spatial dependence of the
ensuing fluctuation spectrum is nevertheless preserved. These findings suggest
that the speed-limited particle-in-cell (SLPIC)
method, as outlined in previous work \cite{Werner:slpic}, is a fast, accurate, and powerful technique
for modeling plasmas wherein electron kinetic behavior is significant (such that a fluid/Boltzmann
representation for electrons is inadequate) but evolution
is on ion timescales.
In these cases the use of PIC is computationally
demanding, but the use of speed-limited electrons can substantially reduce computational
demands without sacrificing the desired physics.

For plasmas with $v_{ti} \ll v_{\phi} \ll v_{te}$ [where these velocities respectively are the ion
thermal velocity, ion acoustic wave phase velocity $\omega/k$ ($\sim c_{s}$ for long-wavelength modes), and the electron thermal velocity],
the speed limit $v_{0}$ can be chosen with $v_{\phi} < v_{0} < v_{te}$.  Choosing $v_{0} < v_{te}$
ensures that the speed-limited plasma oscillation frequency is reduced below the true plasma
oscillation frequency $\omega_{p}$ by a factor $\omega/\omega_{p} \sim v_{0}/v_{te}$, and
allows the timestep to be increased (and the simulation sped up) by a factor $v_{te}/v_{0}$.
However, choosing $v_{0} > v_{\phi}$ ensures that ion acoustic waves are still accurately simulated
(including the Landau damping rate).

Potential applications for the SLPIC method include its use in the modeling of 
plasma thrusters (wherein very
small electron/ion mass ratios impose especially demanding 
numerical constraints) and sheath formation
(e.g.~near a Langmuir probe \cite{wernerlangmuir}).
Collisional low-temperature plasma
discharges are also an area of particular interest;
recent efforts have demonstrated that SLPIC can
be used in conjunction with standard Monte Carlo collision techniques, in the same
manner as is done in collisional PIC discharge modeling (PIC-MCC)
\cite{theis}.  We
anticipate exploring SLPIC's capability for rapid collisional plasma discharge modeling
in future publications.

\begin{acknowledgments}
This research was financially supported by the U.S. Department of Energy, SBIR 
Phase I/II Award DE-SC0015762, and by the U.S. National Science Foundation, 
Grant PHY1707430.  The data that support the findings of this work are available 
from the corresponding author upon reasonable request.  We thank the reviewers 
of this manuscript for their constructive comments.
\end{acknowledgments}

\appendix
\section{Asymptotic expansions of $Z(\gamma,\zeta)$}
\label{appendix:expansions}
A detailed overview of the properties of the incomplete plasma dispersion function 
$Z(\gamma,\zeta)$ was given by Baalrud in Ref.~\citenum{Baalrud:2013}.  A number
of these relations have been used in this work and are summarized here.

For $\zeta \gg 1$:
\begin{equation}
Z(\gamma,\zeta) \sim i \sigma \sqrt{\pi} H(\zeta - \gamma) e^{-\zeta^{2}} 
-{\mbox{erfc}(\gamma) \over 2 \zeta} 
- {e^{-\gamma^{2}} \over 2 \sqrt{\pi} \zeta^{2}} 
- {1 \over \zeta^{3}}\left( {\gamma e^{-\gamma^{2}} \over 2 \sqrt{\pi}} + {\mbox{erfc}(\gamma) \over 4} \right) 
- \ldots
\end{equation}
wherein $H(x)$ is the Heaviside function, $\mbox{erfc}(x)$ is the complementary error function
$\mbox{erfc}(x) = 1 - \mbox{erf}(x)$, and
\begin{equation}
\sigma = 1 - \mbox{sign}[\mbox{Im}(\zeta)] ~.
\end{equation}

For $\zeta \ll 1$:
\begin{equation}
Z(\gamma,\zeta) \sim i \sqrt{\pi} H(\zeta - \gamma) e^{-\zeta^{2}} 
+ {E_{1}(\gamma^{2}) \over 2 \sqrt{\pi}} 
+ \zeta \left( {e^{-\gamma^{2}} \over \gamma \sqrt{\pi}} - \mbox{erfc}(\gamma) \right) 
+ \zeta^{2}\left( {e^{-\gamma^{2}} \over 2 \sqrt{\pi} \gamma^{2}} 
- {E_{1}(\gamma^{2}) \over 2 \sqrt{\pi}} \right) + \ldots
\end{equation}
wherein 
\begin{equation}
E_{1}(x) = \int_{1}^{\infty} {e^{-x t} \over t} ~dt
\end{equation}
is the exponential integral.

Matlab algorithms for evaluating $Z(\zeta)$, $Z(\gamma,\zeta)$, and their derivatives
were also provided in the supplemental materials for Ref.~\citenum{Baalrud:2013}.
These algorithms were used in the numerical calculations of this work.

\section{Constraints and speed-limiting}
\label{appendix:constraints}
In this appendix we briefly consider the scaling of the numerical constraints
outlined in Section \ref{sec:SLPIC} in SLD.

The Debye length resolution constraint is independent of the speed-limiting.  When it
is satisfied, the grid size $\Delta x = \delta \lambda_{De}$ for some $\delta \lesssim 1$.

In SLD, the cell-crossing time constraint is altered by the speed-limiting and becomes
$v_{0} \Delta t < \Delta x$, since no particle can move faster than the speed limit $v_{0}$.
Substituting the result from the Debye length resolution constraint then yields the scaling 
$\Delta t < \delta \lambda_{De}/v_{0}$.

The plasma oscillation constraint, in the limit of aggressive speed-limiting
($\gamma_{e} \rightarrow 0$), replaces $\omega_{p} \sim \omega_{pe}$ by
$\omega_{pe} v_{0}/v_{te}$ (as shown in the small-$\gamma$ limit of Fig. \ref{hfig}).
Substituting the result from the Debye length resolution constraint then yields the scaling 
$\Delta t < 2 v_{te}/(v_{0} \omega_{pe}) = 2 \lambda_{De}/v_{0}$.

It is significant that the timestep $\Delta t$ restriction scales linearly as the ratio of Debye length to
speed limit in both of the latter two constraints.  If this were not so, there could be regions
of parameter space where one constraint or the other prevailed, and the speed-limiting
concept would be less useful.  But the physical scaling for both constraints is the same -- 
the largest timestep for speed-limited particles is on the order of the time required for the 
fastest such particles to cross a Debye length.  This restriction preserves the physics of 
local Debye shielding (a time-independent phenomena) even while slowing the rapid 
plasma oscillations.  In addition, this constraint is independent of the electron-ion 
mass ratio, suggesting that SLPIC can be used even when this ratio is small.

\section{Comparison of speed-limited and relativistic dynamics}
\label{relappendix}
It has been noted that SLD exhibits some similarities with relativistic dynamics, 
wherein the speed of light $c$ plays a role somewhat analogous to the SLD speed limit $v_{0}$.
Although we haven't explored this idea in detail in this work, it is instructive 
to compare the kinetic equation for nonrelativistic SLD [using a
more general Lorentz acceleration term that includes the electromagnetic fields $\mathbf{E}(\mathbf{x},t)$
and $\mathbf{B}(\mathbf{x},t)$] with the relativistic kinetic (Vlasov) equation in the form
\bea
\label{relSLPIC}
   {\partial f_{\alpha} \over \partial t} 
   + \beta \mathbf{v} \cdot {\partial f_{\alpha} \over \partial \mathbf{x}} 
   + \beta {q_{\alpha} \over m_{\alpha}} \mathbf{E} \cdot {\partial f_{\alpha} \over \partial \mathbf{v}} 
   + \beta {q_{\alpha} \over m_{\alpha}} \mathbf{v} \times \mathbf{B} \cdot {\partial f_{\alpha} \over \partial \mathbf{v}}
   & = 0 & ~~~\mbox{(SLD)} \\
   {\partial g_{\alpha} \over \partial t}
   + {1 \over \gamma} \mathbf{u} \cdot {\partial g_{\alpha} \over \partial \mathbf{x}} 
   + {q_{\alpha} \over m_{\alpha}} \mathbf{E} \cdot {\partial g_{\alpha} \over \partial \mathbf{u}} 
   + {1 \over \gamma} {q_{\alpha} \over m_{\alpha}} \mathbf{u} \times \mathbf{B} \cdot {\partial g_{\alpha} \over \partial \mathbf{u}}
   & = 0 & ~~~\mbox{(relativistic)} ~.
\eea
Here, respectively, the SLD distribution $f_{\alpha} = f_{\alpha}(\mathbf{x},\mathbf{v},t)$ is a function of position, velocity, and time, while the relativistic
distribution $g_{\alpha} = g_{\alpha}(\mathbf{x},\mathbf{p},t)$ is a function of position, momentum, and time.  The spatial components of the four-velocity, $\mathbf{u} \equiv (\mathbf{p}/m_{\alpha})$, are related to the conventional three-velocity $\mathbf{v}$ through the relativistic Lorentz factor $\gamma = \sqrt{1+\mathbf{u} \cdot \mathbf{u}/c^2}$, such that $\mathbf{u} = \gamma \mathbf{v}$.

The structure of these equations is very similar.  
The relativistic $\mathbf{u}$ (whose magnitude may exceed $c$) is like the SLD `true velocity' $\mathbf{v}$
(whose magnitude may exceed $v_{0}$), and the factor $1/\gamma$ ($\sim 1$
for $|\mathbf{u}| \ll c$, and $\sim c/|\mathbf{u}|$ for $|\mathbf{u}| \gg c$) plays a role akin
to the speed-limiting function $\beta$ ($\sim 1$ for $|\mathbf{v}| \ll v_0$, and 
$\sim v_{0}/|\mathbf{v}|$ for $|\mathbf{v}| \gg v_0$).  The product of these relativistic functions,
$\mathbf{u}/\gamma$ (three-velocity), can never exceed $c$ just as the SLD `pseudo-velocity' can never exceed $v_{0}$.

Nevertheless, key differences appear.  The term proportional to the electric field, in the relativistic
case, contains no physics equivalent to the speed-limiting that occurs in SLD -- in effect, relativistic physics 
applies the speed-limiting concept to the magnetic, but not the electric, components of the Lorentz 
acceleration.
The ensuing trajectories thus vary from those of SLPIC, wherein the appearance of $\beta$ in all but the first term of Eq.~(\ref{relSLPIC}) can be viewed as a local
rescaling of time (with $\beta$) along a trajectory that is constant regardless of
the value of $v_{0}$.
So while SLD is somewhat like relativistic dynamics, in that it tracks both unbounded (SLD true 
velocity/relativistic momentum) and bounded (SLD pseudo-velocity/relativistic three-velocity) phase space 
variables, with the latter restricted by fixed speed limits (SLD $v_{0}$/relativistic $c$), the dynamics 
of the two systems differ enough to make intuitive comparisons difficult.

\bibliographystyle{phaip}
\bibliography{slpicDispersion}

\newcommand{\SortNoop}[1]{}
\begin{thebibliography}{10}

\bibitem{Werner:slpic}
G.~R. Werner, T.~G. Jenkins, A.~M. Chap, and J.~R. Cary,
\newblock Phys. Plasmas {\bf 25}, 123512 (2018).

\bibitem{Note1}
Subsequent code development has enabled speedup factors of greater than 250 for
  this discharge, relative to conventional PIC. Detailed discharge properties
  are provided in Ref. \protect \citenum {Werner:slpic}.

\bibitem{langdon1970}
A.~B. Langdon and C.~K. Birdsall,
\newblock Phys. Fluids {\bf 13}, 2115 (1970).

\bibitem{Note2}
Collisional effects, which would replace the zero on the right-hand side of Eq.
  (\ref {eq:ConserveVlasov}) with source or sink terms, could also be included
  in this equation; SLPIC is compatible with conventional PIC-MCC techniques
  for modeling collisional plasmas. We will not consider collisional effects in
  this work, but future publications demonstrating collisional SLPIC discharges
  are anticipated.

\bibitem{montgomery}
D.~C. Montgomery, 
\newblock {\em Theory of the Unmagnetized Plasma},
\newblock Gordon and Breach, 1971.
\bibitem{swanson}
D.~G. Swanson,
\newblock {\em Plasma Kinetic Theory},
\newblock CRC Press, 2008.
\bibitem{scheiner2019}
B. Scheiner and P.~J. Adrian,
\newblock Phys. Plasmas {\bf 26}, 034501 (2019).

\bibitem{BirdsallPIC}
C.~K. Birdsall and A.~B. Langdon,
\newblock {\em Plasma Physics via Computer Simulation},
\newblock CRC Press, 2004.

\bibitem{HockneyPIC}
R.~W. Hockney and J.~W. Eastwood,
\newblock {\em Computer Simulation Using Particles},
\newblock CRC Press, 1988.

\bibitem{langdon1970jcp}
A.~B. Langdon,
\newblock J. Comp. Phys. {\bf 6}, 247 (1970).
\bibitem{okuda1970}
H. Okuda and C.~K. Birdsall, 
\newblock Phys. Fluids {\bf 13}, 2123 (1970).
\bibitem{okuda1972}
H. Okuda,
\newblock Phys. Fluids {\bf 15}, 1268 (1972).
\bibitem{langdon1979}
A.~B. Langdon,
\newblock Phys. Fluids {\bf 22}, 163 (1979).

\bibitem{Note3}
Parallels can be drawn with relativistic dynamics, in which
  elements of phase space can move from one physical position to another with
  speed no faster than the speed of light; see Appendix \ref{relappendix} for further discussion.

\bibitem{adam}
J.~C. Adam, A.~Gourdin Serveniere, and A.~B. Langdon,
\newblock J. Comp. Phys. {\bf 47}(2), 229 (1982).

\bibitem{FriedAndConte}
B.~D. Fried and S.~D. Conte,
\newblock {\em The Plasma Dispersion Function},
\newblock Academic Press, 1961.

\bibitem{Franklin}
R.~N. Franklin,
\newblock in {\em Proceedings of the Tenth International Conference on
  Phenomena in Ionized Gases}, page 269, Donald Parsons and Company, Ltd.,
  1971.

\bibitem{Baalrud:2013}
S.~D. Baalrud,
\newblock Phys. Plasmas {\bf 20}, 012118 (2013).

\bibitem{GoldstonTextbook}
R.~J. Goldston and P.~H. Rutherford,
\newblock {\em Introduction To Plasma Physics},
\newblock IoP, 1995.

\bibitem{ChenTextbook}
F.~F. Chen,
\newblock {\em Introduction to Plasma Physics and Controlled Fusion},
\newblock Springer, 3rd edition, 2016.

\bibitem{vorpal}
C. Nieter and J.~R. Cary, 
\newblock J. Comp. Phys. {\bf 196}, 448 (2004).
\bibitem{esirkepov}
T.~Zh. Esirkpov,
\newblock Comp. Phys. Comm. {\bf 135}(2), 144 (2001).

\bibitem{wernerlangmuir}
G. R. Werner, S. Robertson, T. G. Jenkins, A. M. Chap, and J. R. Cary, 
``Accelerated Steady-State Electrostatic Particle-in-Cell Simulation of Langmuir Probes'',
to be submitted to Phys. Plasmas.

\bibitem{theis}
J. Theis, G. R. Werner, T. G. Jenkins, and J. R. Cary, 
Phys. Plasmas {\bf 28}, 063513 (2021).

\end{thebibliography}

\end{document}